 \documentclass[final,5p,times,twocolumn]{elsarticle}

\usepackage{amssymb}
\usepackage{amsfonts}
\usepackage{amsmath}

\usepackage{subfig}
\usepackage{xcolor}

\journal{Physics Letters B}

\begin{document}

\begin{frontmatter}

\title{(Quasi-)normal modes of rotating black holes and new solitons in Einstein-Gauss-Bonnet}

\author[first]{Lilianne Tapia}        
\author[first]{Monserrat Aguayo}
\author[first]{Andr\'{e}s Anabal\'{o}n}
\author[second]{Dumitru Astefanesei}
\author[third]{Nicol\'as Grandi}
\author[fourth]{Fernando Izaurieta}
\author[first]{Julio Oliva}
\author[fourth]{Cristian Quinzacara}

\affiliation[first]{organization={Departamento de Física, Universidad de Concepción},%Department and Organization
	addressline={Casilla, 160-C}, 
	city={Concepción},
        %	postcode={}, 
        %	state={},
	country={Chile.}}

\affiliation[second]{organization={Pontificia Universidad Católica de Valparaíso, Instituto de Física},
	addressline={Av. Brasil 2950},
	city={Valparaíso},
        %	postcode={},
        %	state={},
	country={Chile.}}
	
\affiliation[third]{organization={Instituto de Física La Plata (IFLP), CONICET and Departmento de Física},
	addressline={UNLP C.C. 67, (1900)},
	city={La Plata},
	%	postcode={},
	%	state={},
	country={Argentina.}}

\affiliation[fourth]{organization={Departamento de Ciencias Exactas, Facultad de Ingeniería, Universidad San Sebastián},
        %	addressline={UNLP C.C. 67, (1900)},
        city={Concepción},
	%	postcode={},
	%	state={},
	country={Chile.}}

\begin{abstract}
In this paper, we analyze the scalar field (quasi-)normal modes of recently derived rotating black holes within the framework of Einstein-Gauss-Bonnet theory at the Chern-Simons point in five dimensions. We also examine the mode spectrum of these probes on new static gravitational solitons. These solitons, featuring a regular center, are constructed from static black holes with gravitational hair via a double analytic continuation. By imposing ingoing boundary conditions at the horizons of rotating black holes, ensuring regularity at the soliton centers, and imposing Dirichlet boundary conditions at infinity, we obtain numerical spectra for the rotating black holes and solitons. For static black holes, we demonstrate analytically that the imaginary part of the mode frequencies is negative. Our analysis of the massless Klein-Gordon equation on five-dimensional geometries reveals an infinite family of gapped, massive three-dimensional Klein-Gordon fields, despite the presence of a non-compact extended direction. For the static solitons, the frequencies are real and non-equispaced, whereas in the rotating black holes, counter-rotating modes are absorbed more quickly, and the imaginary part of the co-rotating modes approaches zero as extremality is approached. Additionally, we show that both the rotating black holes and solitons can be equipped with non-trivial torsion, leading to a novel branch of solutions.
\end{abstract}

\end{frontmatter}

%\tableofcontents


\section{Introduction}
\label{sec01}

Recently, a new exact, rotating solution of Einstein-Gauss-Bonnet theory was constructed in dimension five \cite{us} (see \cite{Anabalon:2009kq,Cvetic:2016sow} for other exact rotating solutions at in Lovelock theories at the Chern-Simons point and \cite{rots1}-\cite{rotsfin} for numerical and perturbative solutions in Einstein-Gauss-Bonnet theory). The solution exists at the Chern-Simons point of the theory, with an action principle given by%
\begin{multline}
	I\left[  g_{AB}\right]  =\int d^{5}x\sqrt{-g}\,\bigg[  R-2\Lambda+\\ \alpha\left(
	R^{2}-4R_{AB}R^{AB}+R_{ABCD}R^{ABCD}\right)  \bigg]  \ ,
\end{multline}
where%
\begin{equation}
	\Lambda=-\frac{3}{l^{2}}\text{ and }\alpha=\frac{l^{2}}{4}\ .
	\label{couplings}%
\end{equation}
with field equations
\begin{multline}
	0  =G_{AB}+\Lambda g_{AB}+\\
	\quad\alpha\bigg[  2RR_{AB}-4R_{ACBD}R^{CD}%
	\!+2R_{ACDE}R_{B}^{\ \ CDE}\! -4R_{AC}R_{B}^{\ C}\\
	  -\frac{1}{2}g_{AB}\left(  R^{2}-4R_{AB}%
	R^{AB}+R_{ABCD}R^{ABCD}\right)  \bigg]\ .\label{feqs}%
\end{multline}
(See \cite{Garraffo:2008hu} and \cite{libroCS} and references therein). The five-dimen\-sional geometry, is constructed from a three-dimensional rotating, Kerr-Schild seed with Ricci scalar equal to $-6$, oxidated to dimension five by suitable warp factors \cite{us}. The three-dimensional geometry is asymptotically AdS$_5$, and it is characterized by three parameters $M$, $j$ and $b$, that receive the interpretation of a mass and angular momentum parameters, and a hair of gravitational origin, respectively. The warp factors that allow embedding this geometry as a solution of the five-dimensional theory, contain a further integration constant, $\rho_{0}$, that stands for a second gravitational hair. The metric is integrated assuming a three-dimensional Kerr-Schild ansatz with a single metric function, oxidated to dimension five. After performing the change to Boyer-Lindquist coordinates, the rotating solution reduces to
\begin{multline}
	ds_{\mathrm{BH}_{\mathrm I}}^{2}=l^{2}\cosh^{2}\left(  \rho\right)  \left[  -N^{2}dt^{2}%
	+\frac{dr^{2}}{N^{2}}+r^{2}\left(  d\phi-\frac{j}{2r^{2}}dt\right)
	^{2}\right]\qquad\\
	+l^{2}d\rho^{2}+l^{2}\cosh^{2}\left(  \rho-\rho_{0}\right)
	dx^{2}\ , \label{BLmetric}%
\end{multline}
where%
\begin{equation}
	N^{2}=r^{2}-M-\frac{b}{r}+\frac{j^{2}}{4r^{2}}\text{ }\ . \label{Nder}%
\end{equation}
Here, $l$ is the AdS$_5$ radius of the maximally symmetric solution of the five-dimensional theory, and the ranges of the coordinates are given by $-\infty<t<\infty$, $r_{+}<r<+\infty$, $0\leq\phi<2\pi$, $-\infty<\rho<\infty$ and $-L\leq x\leq L$. The coordinate $x$ can be suitably identified with arbitrary period, or could be considered as well as a non-compact direction. The event horizon is located at $r=r_{+}$, with $r_{+}$, being the largest root of the lapse function $N^{2}\left(  r_{+}\right)  =0$ in \eqref{Nder}. The existence of an event horizon was thoroughly investigated in \cite{us}, and here we will assume that the integration constants are such that the event horizon indeed exists. For simplicity, hereafter we set the second gravitational hair parameter $\rho_{0}$ to zero.

\bigskip

Quasinormal modes of black hole geometries play a fundamental role in our understanding of black hole dynamics. On the one hand, the quasinormal modes dominate the late time, ringdown of black holes that are formed via merger of compact objects (see e.g. \cite{Kokkotas:1999bd},  \cite{Berti:2009kk}). Also, in the context of holography, quasinomal modes of the bulk, asymptotically AdS black hole geometries, capture the relaxation time of thermal perturbations in the boundary theory \cite{Horowitz:1999jd}, \cite{Birmingham:2001pj}, and can be used as well to obtain transport properties of the dual plasma via the computation of the viscosity of the dual system \cite{Kovtun:2005ev}. Given the importance of the physics of quasinormal modes as probe fields on black hole spacetimes, in this paper we study the quasinormal spectrum of the new, exact, rotating black hole geometry given by the metric (\ref{BLmetric}). Even more, we introduce three new spacetimes that solve the equations of Einstein-Gauss-Bonnet theory with a unique vacuum \eqref{feqs}, namely a new rotating black hole, and two static solitons and we explore the propagation of massless scalar probes on these spacetimes.

\bigskip

The paper is organized as follow: Section II contains the characterization of the geometries of the rotating black hole solutions and solitons as well as an asymptotic behavior that accomodates these solutions, and that is preserved by two copies of the generators of the Witt, namely the two dimensional conformal algebra. Section III deals with the problem of normal modes of scalar probes on the solitons, and quasinormal modes for those probes on the rotating black holes. In Section IV we show how to dress our new solutions with non-vanishing, gravitating torsion, which in the context of Einstein-Gauss-Bonnet gravity it is consistent due to the presence of quadratic terms in the curvature. We provide conclusions and further comments in Section V.

\section{A new rotating black hole and two new static solitons}
\label{sec02}

It is known that three-dimensional geometries can be suitably dressed by different families of warp factors to give rise to solutions of Einstein-Gauss-Bonnet with a unique vacuum in dimension five \cite{Giribet:2006ec}, \cite{Dotti:2007az}, \cite{Bogdanos:2009pc}, which can also be extended when branes of codimension one and two are present (see e.g. \cite{Cuadros-Melgar:2007lsd}, \cite{Cuadros-Melgar:2009tdj}, \cite{Cuadros-Melgar:2010ewl}). If we restrict to the case in which the theory can be written as a Chern-Simons theory for the AdS group $SO\left(  4,2\right)$, namely we keep $l\in\mathbb{R}$ in equation (\ref{couplings}), the following spacetime solves the field equations 
\begin{multline}
	\!\!ds_{\mathrm{BH}_{\mathrm{II}}}^{2}=l^{2}\cosh^{2}\left(  \rho\right)  \left[  -N^{2}dt^{2}%
	+\frac{dr^{2}}{N^{2}}+r^{2}\left(  d\phi-\frac{j}{2r^{2}}dt\right)
	^{2}\right]\qquad \\
	 +l^{2}d\rho^{2}+l^{2}e^{2\rho}dx^{2}\ , \label{BLmetric2}%
\end{multline}
with $N\left(  r\right)  $ given as before by equation (\ref{Nder}). As explained in \cite{us}, the crossed term in the metric leading to rotation cannot be removed by a large coordinate transformation when $b\neq0$. Furthermore the asymptotic behavior as $\rho\rightarrow\pm\infty$ in the new rotating black hole (\ref{BLmetric2}) differs from that of (\ref{BLmetric}), since while as $\rho\rightarrow+\infty$ the metric (\ref{BLmetric2}) is asymptotically locally AdS$_{5}$ $\Big(R_{\ \ CD}^{AB}\rightarrow-\frac{1}{l^{2}}\delta_{CD}^{AB}\Big)$, when $\rho\rightarrow-\infty$ the Riemann curvature of (\ref{BLmetric2}) approaches the following components%
\begin{equation}
	R_{\ \ x\rho}^{x\rho}=-\frac{1}{l^{2}}\text{\ ,\ }R_{\ \ \rho\nu}^{\rho\mu
	}=-\frac{1}{l^{2}}\delta_{\nu}^{\mu}\ \text{,\ }R_{\ \ \nu x}^{\mu x}=\frac
	{1}{l^{2}}\delta_{\nu}^{\mu}\ \text{,\ }R_{\ \ \lambda\gamma}^{\mu\nu}%
	=-\frac{1}{l^{2}}\delta_{\lambda\gamma}^{\mu\nu}\ \label{curvaturas},
\end{equation}
where the Greek indices run along the three-dimensional geometry on the square bracket of (\ref{BLmetric2}). Notice that this spacetime is characterized by three integration constants only, the mass and angular momentum parameters $M$ and $j$, and the single gravitational hair $b$.

\bigskip

Both geometries, the rotating black hole (\ref{BLmetric}) and the new one introduced in this work in (\ref{BLmetric2}) can be used to construct static solitons, as follows. Consider the static metric%
\begin{multline}
	\!\!\!\! ds_{\mathrm{BH_{I,II}}}^{2}\!\!=l^{2}\cosh^{2}(\rho)  \Bigg[  -\left(
	r^{2}-M-\frac{b}{r}\right)  dt^{2}+\frac{dr^{2}}{r^{2}-M-\frac{b}{r}}%
	+r^{2}d\phi^{2}\Bigg]\\
		 +l^{2}d\rho^{2}+C^{2}(\rho)\,  dx^{2}\ ,
	\label{bhs}%
\end{multline}
that accommodates both spacetimes (\ref{BLmetric}) and (\ref{BLmetric2}) for suitable choices of the function $C\left(  \rho\right)  $. The double-Wick rotation $\left(  t\rightarrow iy,\phi\rightarrow it\right)  $, leads to%
\begin{multline}
	\!\!\!\!\! ds_{\mathrm{sol_{I,II}}}^{2}\!\!=l^{2}\cosh^{2}(\rho)  \Bigg[  -r^{2}%
	dt^{2}+\frac{dr^{2}}{r^{2}-M-\frac{b}{r}}+\left(  r^{2}-M-\frac{b}{r}\right)
	dy^{2}\Bigg]\\  
	+l^{2}d\rho^{2}+C^{2}(\rho)\, dx^{2}\ .
	\label{solitons}%
\end{multline}
These spacetimes describe smooth gravitational solitons provided the direction parameterized by the coordinate $y$ is compact and has the following period $0\leq y\leq\Delta_{y}$ with
\begin{equation}
	\Delta_{y}=\frac{4\pi r_{0}^{2}}{2r_{0}^{3}+b}\ , \label{periodoy}%
\end{equation}
where $r_{0}$ is the largest root of
\begin{equation}
	r_{0}^{2}-M-\frac{b}{r_{0}}=0\ , \label{errecero}%
\end{equation}
and the range of the radial $r$ coordinate is $\,r_{0}\leq r<+\infty$. The equation (\ref{periodoy}) ensures that the compact direction $\partial_{y}$ shrinks smoothly as $r\rightarrow r_{0}$, avoiding conical defects. One can explicitly check that the two solitons (\ref{solitons}), with $C(\rho)=l\cosh(\rho)$ or $C(\rho)=le^\rho$ fulfill the field equations in (\ref{feqs}). The asymptotic behavior of the solitons (\ref{solitons}) as $\rho\rightarrow\pm\infty$ and $r\rightarrow+\infty$, is
inherited from their static black hole seeds.

Interestingly enough, the rotating black hole solutions we have constructed in \eqref{BLmetric} and \eqref{BLmetric2} are accommodate by the following asymptotic behavior as $r\rightarrow\infty$%
\begin{equation}
	\begin{array}{ll}
		g_{ij}  =\mathcal{O}\left(  1\right)\  ,& g_{i\mu}=\mathcal{O}\left(
		1\right)\ ,  \\[0.5ex]
		g_{rr}   =\dfrac{l^{2}\cosh^{2}(\rho)  }{r^{2}}+\mathcal{O}%
		\left(  r^{-4}\right)  \ ,& g_{tt}=-l^{2}\cosh^{2}(\rho)\,
		r^{2}+\mathcal{O}\left(  1\right) \ , \\[1.2ex]
		g_{tr}   =\mathcal{O}\left(  r^{-3}\right)  \ ,& g_{\phi\phi}=l^{2}\cosh
		^{2}(\rho)\,  r^{2}+\mathcal{O}\left(  1\right)  \ ,\\[1ex] 
		g_{\phi r}=\mathcal{O}\left(  r^{-3}\right)  \ ,&\ g_{t\phi}=\mathcal{O}\left(
		1\right)
	\end{array}
\end{equation}
where $i,j$ stand for $\left\{  \rho,x\right\}  $ and $\mu,\nu$ stand for $\left\{  t,r,\phi\right\}  $. Remarkably, this asymptotic behavior is preserved by the standard infinity dimensional family of vector fields%
\begin{subequations}
	\begin{align}
	\eta^{+}  & =L^{+}\left(  x^{+}\right)  +\frac{l^{2}}{2r^{2}}\partial_{-}%
	^{2}L^{-}\left(  x^{-}\right)  +\mathcal{O}\left(  r^{-4}\right)   \ , \\
	\eta^{r}  &=-\frac{r}{2}\left(  \partial_{+}L^{+}+\partial_{-}L^{-}\right)
	+\mathcal{O}\left(  r^{-1}\right) \ , \\ 
	\eta^{-}&=L^{-}\left(  x^{-}\right)  +\frac{l^{2}}{2r^{2}}\partial_{+}^{2}%
	L^{+}\left(  x^{+}\right)  +\mathcal{O}\left(  r^{-4}\right)  \ ,\\ 
	\eta^{x}&=0\quad ,\qquad \eta^{\rho}=0\ ,
\end{align}
\end{subequations}
where $L^{\pm}\left(  x^{\pm}\right)  $ are arbitrary function of $x^{\pm}=t+l\phi$. These vector fields span two copies the Witt algebra \cite{Brown:1986nw}, which plays a fundamental role on the microscopic description of the entropy of the three-dimensional BTZ black hole \cite{Strominger:1997eq}. It would be interesting to further explore the role of the Witt generators in the context of the sector of Chern-Simons gravity we are considering.

\bigskip

Before moving to the next section, let us emphasize that we have constructed three new solutions of Einstein-Gauss-Bonnet theory at the Chern-Simons point, one of them describing a new rotating black hole (\ref{BLmetric2}), while the latter two correspond to smooth gravitational solitons (\ref{solitons}). In
what follows we will explore the propagation of scalar probes on these geometries.

\section{Scalar probes}
\label{sec03}

Every spacetime described in the previous section (\ref{BLmetric}), (\ref{BLmetric2}) and (\ref{solitons}), are contained within the following family of metrics
\begin{equation}
	d\tilde{s}^{2}=A^{2}(\rho)\,  g_{\mu\nu}dx^{\mu}dx^{\nu}+l^{2}%
	d\rho^{2}+C^{2}(\rho)\,  dx^{2}\ , \label{general}%
\end{equation}
where the tilde $\left(  \symbol{126}\right) $ stands for five-dimensional quantities, and $g_{\mu\nu}$ is a three-dimensional spacetime metric. A massless scalar field%
\begin{equation}
	\square\Phi=0\ , \label{massless}%
\end{equation}
on the geometries (\ref{general}) can be separated as%
\begin{equation}
	\Phi\left(  x^{A}\right)  =\psi\left(  x^{\mu}\right)  Y\left(  \rho\right)
	e^{ikx}\ , \label{FullPhi}%
\end{equation}
leading to the following equations%
\begin{align}
	\partial_{\rho}\left(  A^{3}C\partial_{\rho}Y\left(  \rho\right)  \right)
	+A^{3}Cl^{2}\left(  \frac{m_{\text{eff}}^{2}}{A^{2}}-\frac{k^{2}}{C^{2}}\right)
	Y\left(  \rho\right)   &  =0\ ,\label{eqY}\\
	\square_{3}\psi\left(  x^{\mu}\right)  -m_{\text{eff}}^{2}\psi\left(  x^{\mu}\right)
	&  =0\ . \label{eq3}%
\end{align}
Here $\square_{3}$ is the wave operator on the spacetime metric $g_{\mu\nu}$ in (\ref{general}), and $m_{\text{eff}}^{2}$ emerges as a separation constant, with a clear interpretation from the three-dimensional point of view, namely, that of a mass for the effective three-dimensional scalar. Up to here $m_{\text{eff}}^{2}$ is arbitrary, but it will turn out to be fixed by solving (\ref{eqY}) for suitable boundary conditions for $Y\left( \rho\right)$ as $\rho\rightarrow\pm\infty$. In consequence the problem has been reduced to solving a one-dimensional problem for $Y\left(  \rho\right)$ and a three-dimensional, massive scalar field on the spacetime geometry $g_{\mu\nu}$ that may correspond to that of a rotating black hole or a soliton.

\bigskip

Let us first solve the equation for the $\rho$ dependence. For the rotating black hole and soliton of the Family I in (\ref{BLmetric}) and (\ref{solitons}), respectively, the functions $A\left(\rho\right)  =l\cosh\rho=C\left(  \rho\right)  $ in the general metric (\ref{general}). In this case the equation (\ref{eqY}) admits solutions that vanish as $\rho\rightarrow\pm\infty$. Introducing the change of coordinates $z=\tanh\rho$, the normalizable modes take the form%
\begin{equation}
	Y_{\mathrm I}\left(  \rho\left(  z\right)  \right)  =N_{\mathrm I}\left(  1-z^{2}\right)
	^{2}\ _{2}F_{1}\left(  5+p,-5-p,3,\frac{1-z}{2}\right)  \ ,
\end{equation}
with $p=0,1,2,...$ provided the masses of the Kaluza-Klein tower given by%
\begin{equation}
	m_{\text{eff}}^{2}=m_{\mathrm I}^{2}=k^{2}+\left(  1+p\right)  \left(  4+p\right)  \ ,
	\label{meffI}%
\end{equation}
and $N_{\mathrm  I}$ is a normalization constant. The effective mass, for a given value of $k$, turn
out to take discrete values regardless the fact that the  $\rho$ direction being geometrically non-compact $-\infty<\rho<+\infty$. The negative curvature of the space spanned by the coordinates $\rho$ and $x$ makes the $\rho$ direction to be effectively compact, leading to a gapped, discrete Kaluza-Klein spectrum for the three-dimensional scalars in \eqref{eq3}.

\bigskip

When the rotating black holes and solitons belong to Family II (see equations (\ref{BLmetric2}) and (\ref{solitons})), namely when $A\left(\rho\right) =l\cosh\rho$ and $C\left(  \rho\right)  =le^{\rho}$ in (\ref{general}), the modes that vanish as $\rho\rightarrow\pm\infty$ also exist, in spite of the non-asymptotically locally AdS behavior of the geometry as $\rho\rightarrow-\infty$ as shown explicitly by the curvature components in equation  \eqref{curvaturas}. For generic values of $k$, that controls the dependence of the scalar field on the $x$-coordinates, the equation (\ref{eqY}) is solved in terms of Confluent Heun functions, nevertheless, when $k$ vanishes, the normalizable modes take a simpler form, and are given by%
\begin{equation}
	Y_{\mathrm{II}}\left(  \rho\left(  z\right)  \right)  =A\left(  1-z\right)  ^{2}%
	\ _{2}F_{1}\left(  -p-1,p+3,3,\frac{1-z}{2}\right)
\end{equation}
with $p=0,1,2,...$, implying a discretization of the masses of the KK tower as follows
\begin{equation}
	m_{\text{eff}}^{2}=m_{\mathrm{II}}^{2}=\left(  p+1\right)  \left(  p+3\right)  \ ,
	\label{meffII}%
\end{equation}
In conclusion, we have solved the equation for the dependence of the scalar field in the $\rho$ coordinate, for both families of solutions, that include in turn the solitons and the rotating black holes. Now we move to solving the effective massive, three-dimensional Klein-Gordon problems \eqref{eq3}.

\subsection{Scalar probes on solitons}
\label{subsec0301}

In this section we will obtain the normal frequencies for the massless scalar probe on the soliton geometries (\ref{solitons}). As shown in the previous section, the problem then reduces to solving the massive Klein-Gordon equation%
\begin{equation}
	\square_{3}\psi\left(  x^{\mu}\right)  -m_{\text{eff}}^{2}\psi\left(  x^{\mu}\right)
	=0\ , \label{kg3sol}%
\end{equation}
on the geometry%
\begin{equation}
	ds_{3,\mathrm{sol}}^{2}=-r^{2}dt^{2}+\frac{dr^{2}}{r^{2}-M-\frac{b}{r}}+\left(
	r^{2}-M-\frac{b}{r}\right)  dy^{2}\ , \label{3sol}%
\end{equation}
with $-\infty<t<+\infty$, $r_{0}\leq r<\infty$, and $0\leq y\leq\Delta_{y}$, with $\Delta_{y}$ given in (\ref{periodoy}) and $r_{0}$ fixed as in (\ref{errecero}). The soliton geometry (\ref{3sol}) is asymptotically locally AdS$_{3}$ as $r\rightarrow\infty$, and has a regular center at $r=r_{0}$. It is important to remark that the three dimensional geometry (\ref{3sol}), reduces to global AdS$_{3}$ when $b$ vanishes. This can be explicitly seen by setting $b=0$ in (\ref{3sol}), (\ref{periodoy}) and (\ref{errecero}) and performing the global change of coordinates
\begin{equation}
	t=\frac{T}{r_{0}}\ ,\quad r=r_{0}\sqrt{R^{2}+1}\ ,\quad y=\frac{\psi}{r_{0}}\ ,
	\label{cambio}%
\end{equation}
that maps $ds_{3,\mathrm{sol}}^{2}$ in (\ref{3sol}) to%
\begin{equation}
	ds^{2}=-\left(  R^{2}+1\right)  dT^{2}+\frac{dR^{2}}{R^{2}+1}+R^{2}d\psi
	^{2}\ .
\end{equation}
with $0\leq\phi\leq 2\pi$. In consequence, there is no non-trivial soliton when the hair parameter $b$ vanishes, in spite of the presence of a remaining, seemingly non-trivial constant $r_0$, since the double Wick rotation of static BTZ black hole leads to a spacetime which when regular, it is diffeomorphic to AdS$_{3}$ (see \cite{Horowitz:1998ha}). It is interesting to notice that performing the change of coordinates (\ref{cambio}) in the non-trivial soliton metric (\ref{3sol}) one arrives to the spacetime
metric (setting $R=\sinh z$)%
\begin{multline}
	ds^{2}_{3,\text{sol}}=-\cosh^{2}z\ dT^{2}+\frac{\cosh z\left(  \cosh z+1\right)  %
	}{\tilde{b}+\cosh z\left(  \cosh z+1\right)  }dz^{2}\qquad\\[1ex]
	+\frac{\sinh^{2}z\left(
		\tilde{b}+\cosh z\left(  \cosh z+1\right)  \right)  }{\cosh z\left(  \cosh
		z+1\right)  }d\psi^{2}\, , \label{solcoordnew}%
\end{multline}
where $\tilde{b}=b/r_{0}^{3}$, $0\leq z<+\infty$ and now regularity requires $0\leq\psi\leq\Delta_{\psi}=4\pi/(2+\tilde{b})$. Notice that the soliton (\ref{solcoordnew}) depends on a single integration constant $\tilde{b}$, that parameterizes the departure of the metric from AdS$_{3}$.
Since (\ref{solcoordnew}) depends on non-rational functions of the coordinate $z$, for simplicity we will continue working in the original coordinates (\ref{3sol}). The equation for the radial dependence will turn out to be sensible only to dimensionless ratios $\omega/r_{0}$ and $b/r_{0}^{3}$, consistently with the fact that the spacetime actually depends on a single integration constant.

Assuming the separation for the scalar probe on the hairy soliton (\ref{3sol})%
\begin{equation}
	\psi\left(  x^{\mu}\right)  =e^{-i\omega t+i(2\pi n/\Delta_{y})y%
		}\,R\left(  r\right)  \ ,
\end{equation}
with $n\in\mathbb{Z}$, the Klein-Gordon equation (\ref{kg3sol}) leads to a second-order ODE for $R\left(  r\right)  $, that admits the following asymptotic behavior%
\begin{subequations}
	\begin{align}
	&  R\left(  r\right)  \underset{r\rightarrow r_{0}}{\sim}C_{1}\left(
	r-r_{0}\right)  ^{n/2}\Big(  1+\mathcal{O}\left(  r-r_{0}\right)
	\Big) \nonumber\\
	&\hspace{13.3ex} +C_{2}\left(  r-r_{0}\right)  ^{-n/2}\Big(  1+\mathcal{O}%
	\left(  r-r_{0}\right)  \Big)  \ ,\label{frobsolorigen}\\[1ex]
	&  R\left(  r\right)  \underset{r\rightarrow\infty}{\sim}\frac{D_{1}%
	}{r^{\Delta_{+}}}\bigg(  1+\mathcal{O}\left(  r^{-1}\right)  \bigg)
	+\frac{D_{2}}{r^{\Delta_{-}}}\bigg(  1+\mathcal{O}\left(r^{-1}\right)
	\bigg)  \ , \label{frobsolinf}%
\end{align}
\end{subequations}
where%
\begin{equation}
	\Delta_{\pm}=1\pm\sqrt{1+m_{\text{eff}}^{2}}\ .
\end{equation}

Imposing regularity at the smooth origin $r=r_{0}$, and assuming $n$ positive(negative) requires setting $C_{2}=0$ ($C_{1}=0$) in (\ref{frobsolorigen}). Notice that $m_{\text{eff}}^2$ in (\ref{meffI}) and (\ref{meffII}), are strictly positive, in consequence, for the effective three-dimensional scalar field (\ref{kg3sol}), one is forced to impose Dirichlet boundary conditions as $r\rightarrow\infty$. This, in turn implies that we must set $D_{2}=0$ in order to allow only for normalizable modes. After the boundary conditions are set, one numerically solves the equation for the radial profile interpolating between the physical asymptotic behavior. Such interpolation is possible only for a given set of values of the frequencies $\omega$, that depend on the parameter of soliton $b/r_{0}^{3}$ and on the properties of the scalar mode, namely $\omega=\omega\left(b/r_{0}^{3};m_\text{eff},n,q\right)  $ where $q$ is an integer labeling the overtones. Below we present plots for the frequencies, which are real and therefore lead to normal modes.
\begin{figure*}[h!]
	\centering
	\subfloat[$m_{\text{eff}}(p)$ with $k=0$ for the s-wave $n=0$.]{\includegraphics[scale=1]{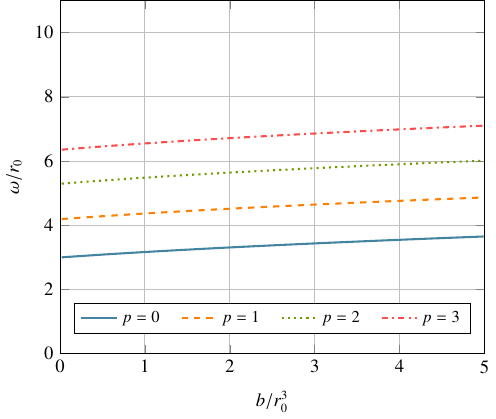}\label{fig01a}}
	\hfill
	\subfloat[$m_{\text{eff}}(p)$ with $k=0$ for $n=1$.]{\includegraphics[scale=1]{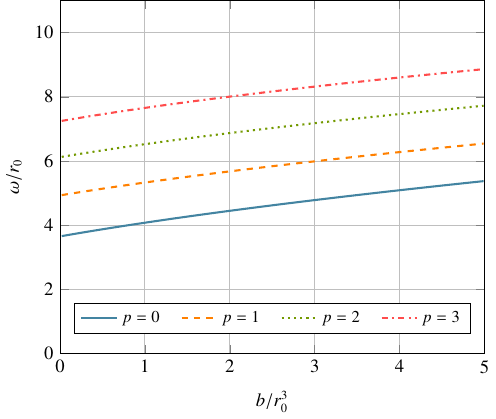}\label{fig01b}}\\[3ex]
	\subfloat[$m_{\text{eff}}(p=0)$ with different values of the angular momentum $n$ of the scalar field.]{\includegraphics[scale=1]{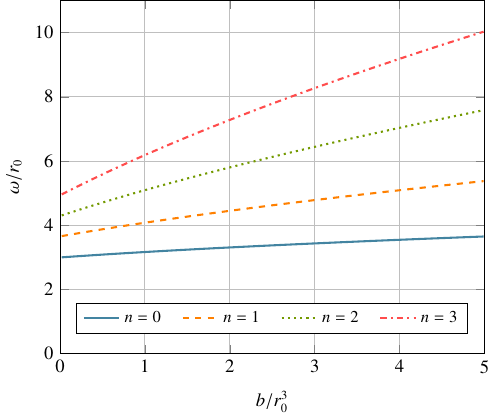}\label{fig01c}}
	\hfill
	\subfloat[ $m_{\text{eff}}(p=1)$ with different values of the angular momentum $n$.]{\includegraphics[scale=1]{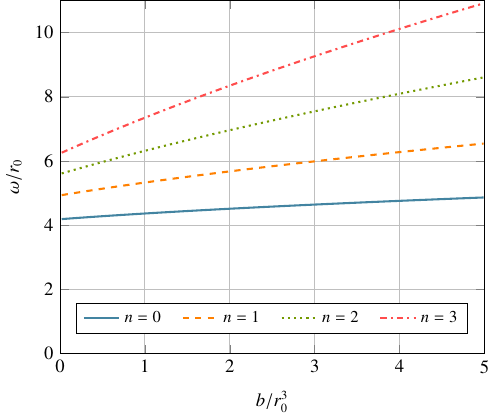}\label{fig01d}}
	\caption{Behavior of the fundamental mode (node-less solution along the radial direction $r$) for the scalar field on the solitons of the Family I, as a function of the parameter $b$. Fig.~\ref{fig01a} corresponds to different values of the effective mass $m_{\text{eff}}(p)$ in \eqref{meffI} with $k=0$ for the s-wave $n=0$, while Fig.~\ref{fig01b} considers the case $n=1$. Fig.~\ref{fig01c} shows the frequencies for the lowest possible value of the effective mass $m_{\text{eff}}(p=0)$ and different values of the angular momentum $n$ of the scalar field. Fig.~\ref{fig01d} corresponds to the same case than the latter but with $m_{\text{eff}}(p=1)$. Exploration of other modes lead to the same qualitative behavior.}
	\label{fig01}
\end{figure*}
The sets $\left(  b/r_{0}^{3},\omega /r_{0}\right)$ and $\left(  r_{0}/b^{1/3},\omega/b^{1/3}\right)$ define
two pairs of dimensionless parameters that allow a thorough understanding of the behavior of the frequencies as the soliton parameter varies. In the former case one is interested in the dependence of the normal frequencies as a function of the gravitational hair $b$ (see Figure \ref{fig01}), this has the advantage of allowing to connect with the spectrum of the scalar probe on global AdS$_{3}$ when $b=0$. In the latter case, when the normal frequencies are plotted on the plane $\left(  r_{0}/b^{1/3},\omega/b^{1/3}\right) $ one explores the behavior of the frequencies as a function of the soliton size (see \ref{fig02}). Notice that the range of the radial coordinate $r$ goes from $r_{0}$ to infinity, and in consequence, in any sensible notion of regularized volume for the soliton, larger values of $r_{0}$ correspond to smaller solitons.
\begin{figure*}[h!]
	\centering
	\subfloat[$m_{\text{eff}}(p)$ with $k=0$ for the s-wave $n=0$.]{\includegraphics[scale=1]{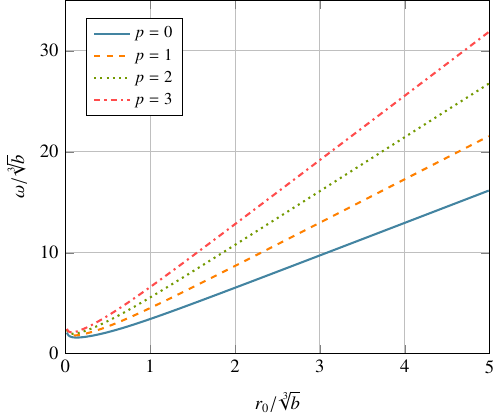}\label{fig02a}}
	\hfill
	\subfloat[$m_{\text{eff}}(p)$ with $k=0$ for $n=1$.]{\includegraphics[scale=1]{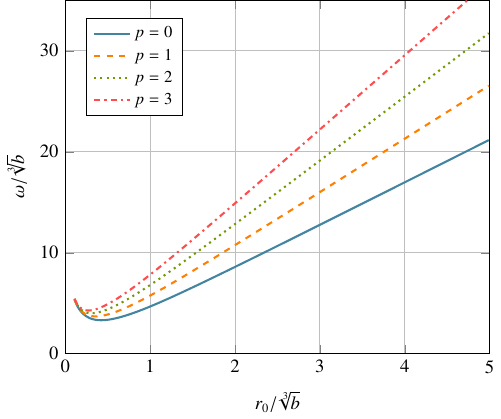}\label{fig02b}}\\[3ex]
	\subfloat[$m_{\text{eff}}(p=0)$ with different values of the angular momentum $n$ of the scalar field.]{\includegraphics[scale=1]{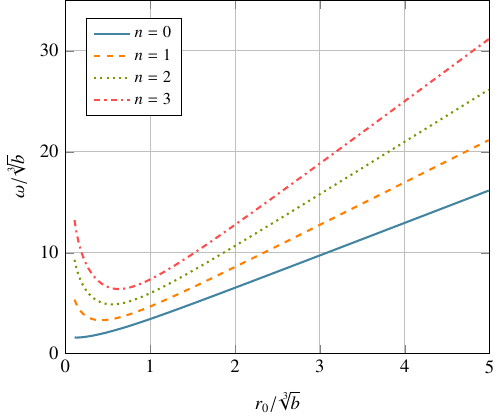}\label{fig02c}}
	\hfill
	\subfloat[ $m_{\text{eff}}(p=1)$ with different values of the angular momentum $n$.]{\includegraphics[scale=1]{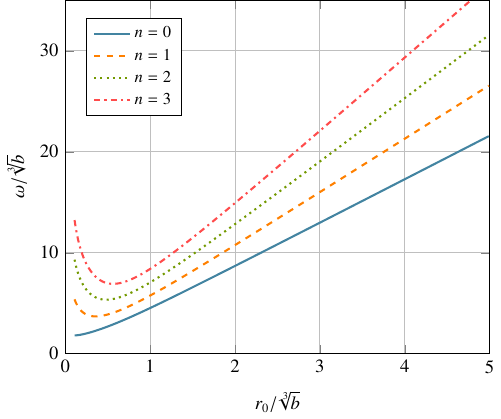}\label{fig02d}}
	\caption{Behavior of the fundamental mode for the scalar field on the Family I solitons as a function of the soliton size $r_0$, for different values of the effective mass $m_{\text{eff}}(p)$ in \eqref{meffI} for $k=0$ and $n=0$ (Fig.~\ref{fig02a}) and $n=1$ (Fig.~\ref{fig02b}), as well as for different values of the angular momentum of the scalar $n$ with fixed effective mass $p=0$ (Fig.~\ref{fig02c}) and $p=1$ (Fig.~\ref{fig02d}).}
	\label{fig02}
\end{figure*}
While the behavior of the fundamental frequency as a function of the parameter $b$ is monotonic, using the soliton size as a control parameter $r_0$, one finds an interesting behavior (see Figure 2). It is expected that for smaller solitons (larger $r_0$) the fundamental frequency must increase, and this is indeed the case for $r_c>r_0$ for a given value of $r_c$ (see \cite{Aguayo:2022ydj} for such monotonic behavior of scalar probes in the context of solitons of $\mathcal{N}=4$ gauged supergravity constructed in \cite{Canfora:2021nca}). Nevertheless, for $r_0<r_c$ the fundamental frequency for the scalar field turns out to be a decreasing function of the soliton size. The critical value of the latter that induces a change on the sign of the slope depends on the mode one is considering.

\subsection{Scalar probes on black holes}

As shown in the previous section, in order to study the behavior of scalar probes on the five dimensional rotating black holes (\ref{BLmetric}) and (\ref{BLmetric2}), one must solve the effective Klein-Gordon equation (\ref{eq3}) on the three-dimensional rotating black hole geometry%
\begin{equation}
	ds_{3,\mathrm{BH}}^{2}=-N^{2}dt^{2}+\frac{dr^{2}}{N^{2}}+r^{2}\left(  d\phi-\frac
	{j}{2r^{2}}dt\right)  ^{2}\ , \label{hairyrot}%
\end{equation}
with $N^{2}=r^{2}-M-\frac{b}{r}+\frac{j^{2}}{4r^{2}}$. Notice that when $b=0$, the spacetime (\ref{hairyrot}) reduces to rotating BTZ black hole. For non-vanishing $b$, it is useful to write the mass parameter $M$ in terms of the horizon location $r_{+}$, namely%
\begin{equation}
	M=\frac{4r_{+}^{4}-4br_{+}+j^{2}}{4r_{+}^{2}}\ .
\end{equation}
The black hole temperature $T$ and the angular velocity of the horizon are given by%
\begin{equation}
	T=\frac{\kappa}{2\pi}=\frac{4r_{+}^{4}+2br_{+}-j^{2}}{8\pi r_{+}^{3}%
	}\quad \text{and}\quad \Omega_{h}=\frac{j}{2r_{+}}, \label{TyOmega}%
\end{equation}
since the horizon generator%
\begin{equation}
	\xi=\partial_{t}+\Omega_{h}\partial_{\phi}\ ,
\end{equation}
fulfils $\xi\cdot\xi\Big\vert _{r_{+}}\!=0$ and $\kappa^{2}=-\frac{1}{2}\xi_{\mu;\nu} \xi^{\mu;\nu}\Big\vert _{r_{+}}$. Assuming a further separation in (\ref{FullPhi})%
\begin{equation}
	\psi\left(  x^{\mu}\right)  =e^{-i\omega t+in\phi}R\left(  r\right)  \ ,
\end{equation}
with $n=0,\pm1,\pm2,...$, in (\ref{eq3}) leads to the following equation for the radial profile $R\left(  r\right)  $ of the scalar probe on the black hole geometries (\ref{BLmetric}) and (\ref{BLmetric2}):%
\begin{multline}
	N^{2}(  r)\,  \frac{d^{2}R}{dr}+\left(  3r-\frac{j^{2}}{4r^{2}%
	}-\frac{M}{r}\right)  \frac{dR}{dr}\\
	-\left(  m_{\text{eff}}^{2}+\frac{n^{2}}{r^{2}%
	}-\frac{\left(  jn-2\omega r^{2}\right)  ^{2}}{4r^{4}N^{2}\left(  r\right)
	}\right)  R=0\ . \label{erre}%
\end{multline}
Notice that the last equation is invariant under the substitution $j\rightarrow-j$, $n\rightarrow-n$, as expected for a rotating spacetime, as well as under the substitution $\omega\rightarrow-\omega$, $j\rightarrow-j$, nonetheless the latter is broken by the quasinormal mode boundary conditions, as shown below. For non-vanishing $b$, this equation must be integrated numerically. Nevertheless, for the static case, in spite of the non-integrability of this equations, when $M\geq 0$ one can prove that for quasinormal mode boundary conditions at the horizon, and Dirichlet boundary conditions at infinity, the imaginary part of the frequency is negative, leading to a stable propagation on the static black hole background. Following the argument developed in \cite{Horowitz:1999jd}, let us first consider the behavior of the function $R(r)$ near the horizon $r=r_+$
\begin{multline}
	R\left(  r\right)  =B_{1}\left(  r-r_{+}\right)  ^{-(\omega-\Omega_{h}n)/(4\pi T)\,i}\left(  1+\mathcal{O}\left(	r-r_{+}\right)  \right) \\
	 +B_{2}\left(  r-r_{+}\right)  ^{(\omega-\Omega_{h}n)/(4\pi T)\,i}\left(  1+\mathcal{O}\left(
	r-r_{+}\right)  \right)  \ , \label{nearhorizon}%
\end{multline}
where the temperature $T$ and the rotation velocity of the horizon $\Omega_{h}$ are given in (\ref{TyOmega}). The quasinormal mode boundary condition at the horizon corresponds to choosing $B_{2}=0$ in (\ref{nearhorizon}). When $j=0$, the three-dimensional, static black hole metric can be
written as%
\begin{equation}
	ds^{2}_{3,\text{BH},j=0}=-f\left(  r\right)  dv^{2}+2dudr+r^{2}d\phi^{2}\ ,
\end{equation}
with $f\left(  r\right)  =r^{2}-M-b/r$, and assuming the separation for the effectively massive, three-dimensional Klein-Gordon field
\begin{equation}
	\psi=\frac{P\left(  r\right)  }{r^{1/2}}e^{-i\omega v+in\phi}\ ,
\end{equation}
leads to the equation
\begin{equation}
	f\left(  r\right)  P^{\prime\prime}+\left(  f^{\prime}-2i\omega\right)
	P^{\prime}-V\left(  r\right)  P=0\ , \label{eqHH}%
\end{equation}
with%
\begin{equation}
	V(r)=\frac{1}{4r^{2}}\left(  4m_{\text{eff}}^{2}r^{2}+4n^{2}+2f^{\prime}r-f\right)
	\ . \label{potHH}%
\end{equation}
One can prove that the potential $V(r)$ in (41) is strictly positive for $r\geq r_+$, provided $M \geq 0$, regardless of the value of $b$. Notice that these black hole geometries are continuously connected with BTZ black holes as $b$ approaches zero. Therefore, manipulating (\ref{eqHH}) and imposing analyticity of the quasinormal modes at the horizon and Dirichlet boundary conditions at infinity, one obtains%
\begin{equation}
	\int_{r_{+}}^{\infty}dr\left(  f\left(  r\right)  |P^{\prime}|^{2}+V\left(
	r\right)  |P|^{2}\right)  =-\frac{|\omega|^{2}|P\left(  r_{+}\right)  |^{2}%
	}{\operatorname{Im}\left(  \omega\right)  }\ .
\end{equation}
In consequence, the positivity of $V\left(  r\right) $ implies that $\operatorname{Im}\left(  \omega\right)  <0$, and then the scalar field propagates in a stable manner on the static black hole.

The previous argument does not apply when $b\neq0$ in the rotating case, nevertheless since $b=0$ leads to a three-dimensional rotating BTZ geometry, such case can treated in analytic fashion. Indeed, quasinormal modes with Dirichlet boundary condition at infinity were originaly computed in \cite{Cardoso:2001hn}, \cite{Birmingham:2001hc} and are given by
\begin{align}
	\omega_{n_{1}}^{\left(  b=0\right)  }  &  =-n-\frac{\left(  2r_{+}%
		^{2}+j\right)  }{2r_{+}}\left(  2n_{1}+1+\sqrt{1+m_{\text{eff}}^{2}}\right)
	i\ ,\label{freq1}\\
	\omega_{n_{2}}^{\left(  b=0\right)  }  &  =n-\frac{\left(  2r_{+}%
		^{2}-j\right)  }{2r_{+}}\left(  2n_{2}+1+\sqrt{1+m_{\text{eff}}^{2}}\right)  i\ ,
	\label{frea2}%
\end{align}
where the mode numbers $n_{1}$ and $n_{2}$ belong to $\left\{0,1,2,...\right\}  $. Notice that the frequencies are mapped into each other if we interchange $n\rightarrow-n$, and $j\rightarrow-j$, namely if we simultaneously flip the direction of the rotation of the black hole background, together with the direction of the angular momentum of the scalar. The imaginary part of the frequencies for these modes is always negative, and therefore the scalar field is being absorbed by the black hole (the existence of a horizon in the rotating BTZ geometry requires $2r_+^2\geq j$ the equality being achieved at extremality). Since $m_{\text{eff}}^{2}$ in (\ref{meffI}) and (\ref{meffII}) are strictly positive, it would be impossible to trigger a superradiant instability choosing Robin boundary conditions as done \cite{Dappiaggi:2017pbe}, since normalizability in our case requires Dirichlet boundary conditions. In consequence, without lost of generality restricting $j\geq0$, one has that the modes $\omega_{n_{1}}^{\left(  b=0\right) }$ in \eqref{freq1} correspond to modes that counter-rotate with respect to the black hole since those modes lead to a phase of the form $e^{in(t+\phi)}$, while the modes $\omega_{n_{1}}^{\left(  b=0\right) }$ in \eqref{frea2} co-rotate with the black hole (they are proportional to $e^{-in(t-\phi)}$). Since the imaginary part of both sets of frequencies are different, co- and counter-rotating modes will have different lifetimes. Simple inspections shows that for $n_2=n_1$, namely for modes with the same number of nodes along the radial direction one has
\begin{equation}
	\operatorname{Im}\left(\omega_{n_{2}}^{\left(  b=0\right)  }\right)<\operatorname{Im}\left(\omega_{n_{1}}^{\left(  b=0\right)  }\right)\ ,
\end{equation}
therefore the co-rotating modes live longer than the counter-rotating ones.

In what follows, we focus on the computation of the quasinormal frequencies for $b\neq0$, considering $b$ as a parameter inducing a departure from BTZ geometry, namely we will focus on the half-plane $M\geq|j|$ (notice that the
presence of $b$ allows for the existence of black hole horizons even outside this region \cite{us}).

\begin{figure*}[h!]
	\centering
	\subfloat[Real part of the quasinormal frequencies.]{\includegraphics[scale=1]{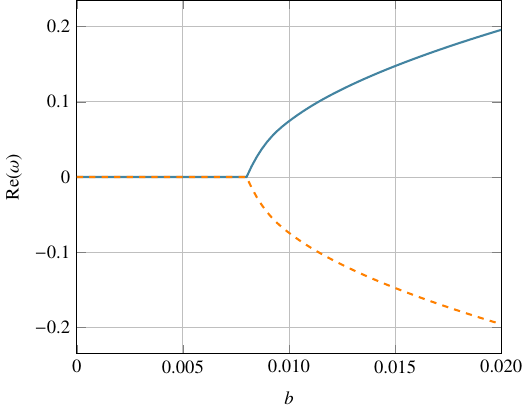}\label{fig03a}}
	\hfill
	\subfloat[Imaginary part of the quasinormal frequencies.]{\includegraphics[scale=1]{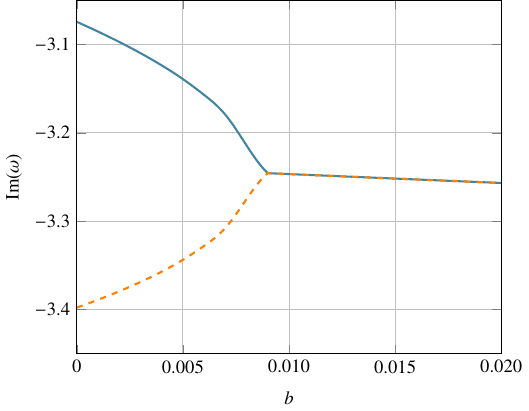}\label{fig03b}}
	\caption{Real (Fig.~\ref{fig03a}) and imaginary (Fig.~\ref{fig03b}) parts of the quasinormal frequencies for the rotating black hole with $j=1/10$, $r_+=1$ and for different values of $b$, for the s-wave scalar field and $m_{\text{eff}}=4$.}
	\label{fig03}
\end{figure*}

\begin{figure*}[h!]
	\centering
	\subfloat[Counter-rotating modes.]{\includegraphics[scale=1]{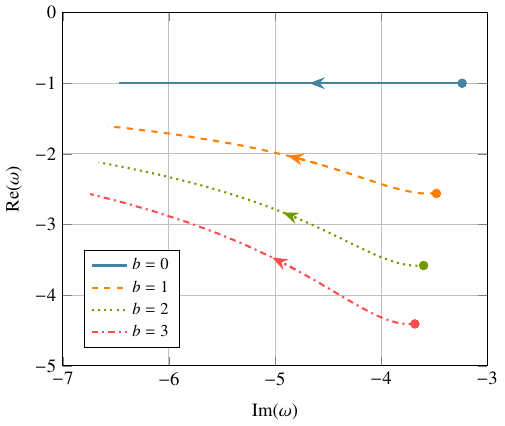}\label{fig04a}}
	\hfill
	\subfloat[Co-rotating modes.]{\includegraphics[scale=1]{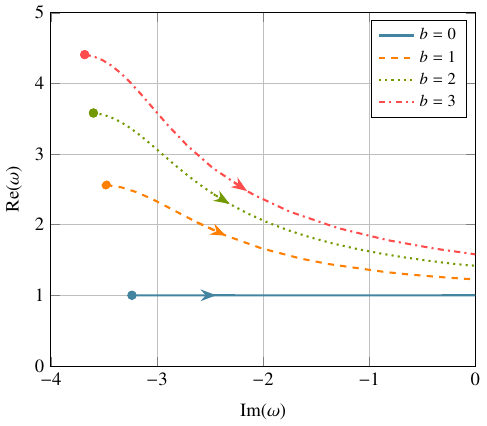}\label{fig04b}}
	\caption{Different curves correspond to different values of the parameter $b$, for the fundamental modes (lowest harmonic number). Here the angular momentum of the scalar field has been fixed as $n=1$. The curves start on the dots, and the arrows indicate the direction of growth of the angular momentum $j$. For $b=0$ the curves are given by \eqref{freq1} ($\operatorname{Re}(\omega)<0$ counter-rotating modes) and \eqref{frea2} ($\operatorname{Re}(\omega)>0$ co-rotating modes). As the angular momentum of the black hole increases, the imaginary part of the counter-rotating modes becomes smaller and the modes are absorbed at a higher rate. On the other hand for the co-rotating modes, as the angular momentum of the black hole increases the imaginary part of the frequencies increase and approaches zero at extremality.}
	\label{fig04}
\end{figure*}

In order to produce the numerical results we have used a standard power series solutions intertwined with Runge-Kutta numerical integration routines built-in in Mathematica. We have also used the QNMSpectral package developed by A. Jansen in \cite{Jansen:2017oag} as a cross-check to explore regions of the parameter space with slower convergence. Even though in the latter reference the QNMSpectral code is applied only to static backgrounds, it turns out to work perfectly in our rotating, asymptotically AdS, hairy solutions. In order to use the QNMSpectral package in an optimal manner, it is suggested to work in ingoing coordinates. The reason is two-folded: first, the QNM boundary condition at the horizon, turns out to be imposing regularity of the scalar field at the horizon, while the second reason is given by the fact that the differential eigenvalue problem to be solved, implies a first order algebraic equation on the frequencies. There are no $\omega^2$ terms since in ingoing coordinates $g^{vv}=0$. Our effective, three-dimensional, rotating black hole \eqref{hairyrot} after performing the transformation 
\begin{equation}
	dv=dt+\frac{dr}{N(r)^2}\ , \quad d\tilde{\phi}=d\phi+\frac{j}{2r^2N(r)^2}\ ,
\end{equation}
leads to the rotating black hole in ingoing Eddington-Fin\-kel\-stein coordinates, namely the metric reads
\begin{equation}
	ds^2=-N(r)^2dv^2+2dvdr+r^2\left(d\tilde{\phi}-\frac{j}{2r^2}dv\right)^2\ ,
\end{equation}
with $N^{2}=r^{2}-M-b/r+j^{2}/(4r^{2})$, and on this spacetime one considers the separation for the effective three-dimensional massive scalar, as follows
\begin{equation}
	\psi\left(v,r,\tilde{\phi}\right)=e^{-i\omega v}J(r)e^{i n \tilde{\phi}}
\end{equation}
The spectra obtained are presented in Figure~\ref{fig03} for the s-wave modes, namely for $n=0$ and in Figure~\ref{fig04} for $n=1$.

For the fundamental mode, s-wave scalar field, the behavior of the frequencies is very intricate, as shown in Figure~\ref{fig03}. We have focused on the case with small angular momentum of the rotating hole $j=1/10$ and $r_+=1$. For vanishing $b$, the frequencies are purely imaginary, and are consistent with the expressions for the BTZ black hole in \eqref{freq1} and \eqref{frea2}, leading to two different, purely imaginary values when the mode numbers $n_1=0=n_2$ vanish, namely for the fundamental mode. As $b$ increases, the frequencies remain purely imaginary, but the two possible values for $\operatorname{Im}(\omega)$ converge to the same number for a critical value of the hair parameter $b\sim 0.08$ above which the imaginary part of both modes coincide, but the real part branches out in co-rotating and counter-rotating modes.   

For the fundamental mode but with rotation on the scalar field, namely with $n=1$ the situation is different. The spectra for such case is depicted in Figure~\ref{fig04}, and interpreted in the caption of such figure.

\section{Dressing the rotating black holes and solitons with torsion in vacuum}
\label{sec04}

It is known that the first order formulation of higher curvature gravities may lead to a non-trivial torsion (see e.g. \cite{Canfora:2007ux,Canfora:2007xs} and the more recent \cite{Barrientos:2017utp,Alvarez:2022wcj}). For Einstein-Gauss-Bonnet at the Chern-Simons point, this is indeed the case. Maintaining the attention on the case with negative cosmological constant, as mentioned above the EGB theory when $\Lambda\alpha=-\frac{3}{4}$ in five dimensions can be formulated
as a Chern-Simons theory. In this section we show that it is possible to dress the rotating black hole spacetime (\ref{BLmetric}) with a non-vanishing torsion, paving the path to potential embeddings of these solutions in Chern-Simons supergravity \cite{izau,Giribet:2014hpa,Andrianopoli:2021qli}.

It will prove useful to rewrite the action functional in terms of the vielbein $e^{A}$, and the curvature 2-form $R^{AB}$, which in terms of the Lorentz connection $\Omega^{AB}$ is given by $R^{AB}=d\Omega^{AB}+\Omega_{\ C}^{A}\Omega^{CB}$, and $d$ is the exterior derivative. The action takes the form%
\begin{equation}
	I\left[  e,\Omega\right]\!  =\!\kappa\!\int\epsilon_{ABCDE}\bigg(  R^{AB}%
	R^{CD}\!+\frac{2}{3l^{2}}R^{AB}e^{C}e^{D}+\frac{1}{5l^{4}}e^{A}e^{B}e^{C}%
	e^{D}\bigg) \, e^{E}\ , \label{csaction}%
\end{equation}
where $\kappa$ is a dimensionful constant, $\epsilon_{ABCDE}$ is the Levi-Civita tensor invariant under $SO\left(  4,1\right)  $. Exterior wedge products $\wedge$ are understood between $p-$forms, and will be explicitly given in some cases to avoid ambiguities. In this section, capital Latin indices stand for $SO\left(  4,1\right)  $ local Lorentz indices. The field equations obtained from (\ref{csaction}) by taking variations with respect to the vielbein $e^{A}$ and the spin-connection $\Omega^{AB}$, are respectively given by%
\begin{subequations}
	\begin{equation}\label{epsA}
		\begin{split}
		\varepsilon_{A}  &=\epsilon_{ABCDE}\bar{R}%
		^{BC}\bar{R}^{DE}\\
		&=\epsilon_{ABCDE}R^{BC}R^{DE}+\frac{2}{l^{2}}R^{BC}%
		e^{D}e^{E}+\frac{1}{l^{4}}e^{B}e^{C}e^{D}e^{E}=0\ ,
		\end{split}
	\end{equation}
	\begin{equation}
	\varepsilon_{AB}   =\epsilon_{ABCDE}\bar{R}^{CD}T^{E}=\epsilon_{ABCDE}\left(  R^{CD}+\frac{1}{l^{2}}e^{C}%
	e^{D}\right)  T^{E}=0 \label{epsAB}%
	\end{equation}
\end{subequations}
where%
\begin{equation}\label{errebarra}
	\bar{R}^{AB}=R^{AB}+\frac{1}{l^{2}}e^{A}e^{B}\ ,
\end{equation}
and the torsion 2-form is given by $T^{A}=De^{A}=de^{A}+\Omega_{\ \ B}^{A}e^{B}$. Here $D$ stands for the covariant derivative under local Lorentz transformations in $SO\left(  4,1\right)  $ with respect to the torsionful spin connection $\Omega_{\ \ B}^{A}$. We will be interested in solution of the
five-dimensional field equation (\ref{epsA})-(\ref{epsAB}) with non-vanishing torsion, within the ansatz%
\begin{equation}
	ds^{2}=A^{2}\left(  \rho\right)  \tilde{g}_{\mu\nu}dx^{\mu}dx^{\nu}+B\left(
	\rho\right)  ^{2}d\rho^{2}+C\left(  \rho\right)  ^{2}dx^{2}\ ,
	\label{metricgen}%
\end{equation}
where the metric $\tilde{g}_{\mu\nu}$ stands for the metric on the three-dimensional rotating spacetime
\begin{equation}
	\begin{split}
	d\tilde{s}^{2}&=\tilde{g}_{\mu\nu}dx^{\mu}dx^{\nu}\\
	&=-\left(  r^{2}-M-\frac{b}%
	{r}+\frac{j^{2}}{4r^{2}}\right)  dt^{2}+\frac{dr^{2}}{r^{2}-M-\frac{b}%
		{r}+\frac{j^{2}}{4r^{2}}}\\
		&\hspace{28ex}+r^{2}\left(  d\phi-\frac{j}{2r^{2}}dt\right)
	^{2}\ . \label{trestilde}%
\end{split}
\end{equation}
Notice that in order to simplify the notation when using Ein\-stein-Cartan geometry, in the present section we have adopted a different notation than in the rest of the paper. For the spacetime (\ref{metricgen}), we can choose the vielbein components as $e^{A}=\left\{  e^{a},e^{4},e^{5}\right\}$ with%
\begin{equation}
	e^{a}=A\left(  \rho\right)  \tilde{e}^{a}\ ,\quad e^{4}=B\left(  \rho\right)
	d\rho\ ,\quad e^{5}=C\left(  \rho\right)  dx\ , \label{viel5}%
\end{equation}
where we split the five dimensional local Lorentz indices $A\in\left\{  a,4,5\right\}  $, and lower-case Latin indices stand for Lorentz indices on three-dimensional spacetime with metric $\tilde{g}_{\mu\nu}$, namelly $\tilde{e}_{\ \mu}^{a}\tilde{e}_{\ \nu}^{b}\eta_{ab}=\tilde{g}_{\mu\nu}$ and $\eta_{ab}$ denoting the Lorentz invariant, three-dimensional Minkowski metric $\eta_{ab}=diag\left(  -1,1,1\right)$. Since it has been proven useful in the literature \cite{Canfora:2007ux,Canfora:2007xs}, we will dress the spacetime (\ref{metricgen}) with the torsion%
\begin{equation}
	T_{a}=-H\left(  \rho\right)  \epsilon_{abc}e^{b}e^{c}\ ,
\end{equation}
that comes from the contorsion%
\begin{equation}
	k_{ab}=HA\epsilon_{abc}\tilde{e}^{c}\ , \label{contorsion}%
\end{equation}
since $T^{A}=De^{A}=k_{\ B}^{A}e^{B}$. Here $\epsilon_{abc}$ is the three-dimensional $SO\left(2,1\right)$ invariant Levi-Civita tensor. Considering the contorsion (\ref{contorsion}) and the vielbein (\ref{metricgen}), the components of the torsionful spin connection $\Omega^{AB}=\mathring{\Omega}^{AB}+k^{AB}$ are given in terms of the torsionless spin-connection $\mathring{\Omega}^{AB}$ and read%
\begin{subequations}
	\begin{gather}
	\Omega^{ab}=\mathring{\omega}^{ab}+k^{ab}\ , \quad \Omega^{45}=\mathring
	{\Omega}^{45}=-\frac{C^{\prime}}{B}dx\ ,\\ 
	\Omega^{m4}=\mathring{\Omega}^{m4}=\frac{A^{\prime}}{B}\tilde{e}^{m}\ .
\end{gather}
\end{subequations}
Here $\mathring{\omega}^{ab}$ is the torsionless spin-connection on the three-dimen\-sional spacetime with metric $\tilde{g}_{\mu\nu}$, namely $d\tilde{e}^{a}+\mathring{\omega}_{\ b}^{a}\tilde{e}^{b}=0$. The components of the curvature two-form $R^{AB}=\mathring{R}^{AB}+\mathring{D}k^{AB}+k_{\ C}^{A}k^{CB}$, where $\mathring{D}$ is the Lorentz covariant derivative with respect to the torsionless connection $\mathring{\Omega}^{AB}$, are given
by%
\begin{subequations}
	\begin{gather}
	R^{ab}  =\mathring{R}^{ab}-\left(  \frac{A^{\prime}}{B}\right)  ^{2}%
	\tilde{e}^{a}\tilde{e}^{b}+d\left(  HA\right)  \epsilon^{abc}\tilde{e}%
	_{c}-\left(  HA\right)  ^{2}\tilde{e}^{a}\tilde{e}^{b}\ ,\\ 
	R^{a4} =\left(  \frac{A^{\prime}}{B}\right)  ^{\prime}d\rho\tilde{e}%
	^{a}-\frac{AA^{\prime}H}{B}\epsilon^{abc}\tilde{e}_{b}\tilde{e}_{c}%
	\ ,\\
	R^{45}=-\left(
	\frac{C^{\prime}}{B}\right)  ^{\prime}d\rho\wedge dx\quad ,\qquad R^{a5}=-\frac{A^{\prime}C^{\prime}}{B^{2}}\tilde{e}^{a}dx\ ,
\end{gather}
\end{subequations}
where $\mathring{R}^{ab}$ is the Riemann curvature two-form of $\tilde{g}_{\mu\nu}$ in (\ref{trestilde}). Let us assume $H\left(  \rho\right)  A\left(\rho\right)  =\delta$ with $\delta$ a constant, then the components of $\bar{R}^{AB}$ defined in \eqref{errebarra} are%
\begin{subequations}
	\begin{align}
	\bar{R}^{ab}  &  =\mathring{R}^{ab}+\left(  \frac{A^{2}}{l^{2}}-\delta
	^{2}-\left(  \frac{A^{\prime}}{B}\right)  ^{2}\right)  \tilde{e}^{a}\tilde
	{e}^{b}\ ,\\ 
	\bar{R}^{45}&=\left(  -\left(  \frac{C^{\prime}}{B}\right)  ^{\prime
	}+\frac{BC}{l^{2}}\right)  d\rho dx\ ,\\
	\bar{R}^{a4}  &  =\left(  \left(  \frac{A^{\prime}}{B}\right)  ^{\prime}%
	-\frac{AB}{l^{2}}\right)  d\rho\tilde{e}^{a}-\frac{A^{\prime}\delta}%
	{B}\epsilon^{abc}\tilde{e}_{b}\tilde{e}_{c}\ ,\\ 
	\bar{R}^{a5}&=\left(
	-\frac{A^{\prime}C^{\prime}}{B^{2}}+\frac{AC}{l^{2}}\right)  \tilde{e}^{a}dx,
\end{align}
\end{subequations}
The equations $\varepsilon_{AB}=0$ in (\ref{epsAB}) strongly constrain the unknown functions, which for non-vanishing $\delta$ lead to%
\begin{subequations}\label{ecs6466}
	\begin{align}
	\varepsilon_{4a}    =0\rightarrow\ -\frac{A^{\prime}C^{\prime}}{B^{2}}%
	+\frac{AC}{l^{2}}&=0\rightarrow\bar{R}^{5d}=0\ ,\label{s1}\\
	\varepsilon_{5a}    =0\rightarrow\,\ \ \ \left(  \frac{A^{\prime}}{B}\right)
	^{\prime}-\frac{AB}{l^{2}}&=0\rightarrow\bar{R}^{a4}=-\frac{A^{\prime}\delta
	}{B}\epsilon^{abc}\tilde{e}_{b}\tilde{e}_{c}\ ,\label{s2}\\
	\varepsilon_{ab}    =0\rightarrow-\left(  \frac{C^{\prime}}{B}\right)
	^{\prime}+\frac{BC}{l^{2}}&=0\rightarrow\bar{R}^{45}=0\ . \label{s3}%
\end{align}
\end{subequations}
The equation $\varepsilon_{45}=0$ is identically fulfilled. Hereafter we fix the gauge $B\left(\rho\right)=l$. The system (\ref{ecs6466}) finally implies 
\begin{subequations}
	\begin{align}
	A\left(  \rho\right)   &  =C_{1}\cosh\left(  \rho\right)  +C_{2}\sinh\left(
	\rho\right) \\
	C\left(  \rho\right)   &  =C_{2}\cosh\left(  \rho\right)  +C_{1}\sinh\left(
	\rho\right)
\end{align}
\end{subequations}
The equations $\varepsilon_{A}=0$ in (\ref{epsA}) are then identically fulfilled. Here, $C_{1}$ and $C_{2}$ are integration constants. In order to avoid curvature singularities, and to preserve an AdS asymptotic behavior as $\rho\rightarrow\infty$ we fix them such that $A(\rho) =l\cosh\left(  \rho-\rho_{0}\right)  $ and $C(\rho) =l\sinh\left(\rho-\rho_{0}\right)  $, but $\rho_{0}$ can be removed by a shift in $\rho$. It can be checked that this choice indeed leads to a spacetime devoid of curvature singularities, except the one at $r=0$.

In summary, the physically sensible spacetimes finally read%
\begin{multline}
	ds^{2}=l^{2}\cosh^{2}\rho\left[  -N^{2}dt^{2}+\frac{dr^{2}}{N^{2}}%
	+r^{2}\left(  d\phi-\frac{j}{2r^{2}}dt\right)  ^{2}\right]\\
	+l^{2}d\rho	^{2}+l^{2}\sinh^{2}\rho dx^{2}\ ,
\end{multline}
with $N^{2}=r^{2}-M-\frac{b}{r}+\frac{j^{2}}{4r^{2}}$, dressed by the torsion%
\begin{equation}\label{laTfinal}
	T_{a}=-l\delta\cosh\rho\epsilon_{abc}\tilde{e}^{b}\tilde{e}^{c}\ ,
\end{equation}
where $\tilde{e}^{b}$ is the dreibein of (\ref{trestilde}).

The torsionful solution is characterized by four integration constant $M$, $j$, $b$ and $\delta$. Notice that this solution is disconnected from the branches discussed in the previous sections since when the torsion vanishes, namely when $\delta=0$, the torsionful solution does not reduce to those discussed above. When the angular momentum parameter $j$ vanishes, one can construct torsionful gravitational solitons, by suitably identifying the vierbein component in \eqref{laTfinal}, after performing the double Wick rotation on the metric. 

\section{Conclusions}

In this work we have constructed new rotating black holes and static solitons of Einstein-Gauss-Bonnet theory at the Chern-Simons point, and studied the propagation of a massless scalar probe on these geometries. The backgrounds are locally analogous to the solutions of the form $F(x)ds^2_{\text{KerrAdS}}+d\Sigma(x)^2_{H_2}$ of General Relativity with a negative cosmological constant, where $F(x)$ is a suitable warp-factor, $ds^2_{\text{KerrAdS}}$ is the Kerr-AdS rotating black hole solution, and $d\Sigma(x)^2_{H_2}$ is the metric of a two-dimensional hyperbolic space. In consequence, they can be interpreted as curved black branes that contain a deformed version of the rotating BTZ black hole on the three-dimensional brane, or a soliton. Remarkably, the rotating solutions can be accommodated in an asymptotic behavior that is preserved by two copies of the Witt algebra. For the rotating black holes the scalar propagation is determined by quasi-normal frequencies that are sensitive to the angular momentum, the mass and a hair of gravitational origin that parameterize the geometry. On the other hand, on the static soliton, the frequencies depend on the single integration constant that characterizes the geometry. The problem of solving the five dimensional scalar probe reduces to that of solving an ODE along the $\rho$ direction, that connects two asymptotic regions $\rho\rightarrow\pm\infty$, that may or may not be locally AdS. On top of such equation, one must solve an effective three-dimensional, massive Klein-Gordon equation. The masses are determined by the quantum numbers of the scalar on the directions orthogonal to the three-dimensional metric, and in spite of the fact that such space is non-compact, the spectrum of masses is discrete and gapped. For the rotating black holes, we have shown that the fundamental modes with co- and counter- rotation with respect to the black hole, behave in an striking different manner as the angular momentum of the black hole increases. While the counter-rotating modes acquire a larger damping as the angular momentum of the rotating hole increases, the co-rotating modes live longer as the angular momentum increases, approaching the real axis as the background approaches extremality.

The interpretation of our black hole solutions as curved black branes mentioned above, suggests that gravitational perturbations may trigger a Gregory-Laflamme instability \cite{GL1,GL2} (see also \cite{GL3}). Let us first mention that for generic values of the coupling $\alpha$ the stability of asymptotically flat and AdS black holes in Einstein-Gauss-Bonnet gravity has been studied in the series of works \cite{DG1,DG2} and \cite{KZ1,KZ2,KZ3}, for the Boulware-Deser black hole \cite{Boulware:1985wk} and its topological generalizations\footnote{The literature on quasinormal modes of black holes in Einstein-Gauss-Bonnet theory is extensive and it is beyond the scope of this paper to provide a full bibliographic review of the topic, nevertheless, in the case of asymptotically locally AdS black holes, refer e.g. to references \cite{QNM0} and \cite{QNM1}, and references therein and thereof.}. The key point is that such analysis of gravitational perturbations can be consistently performed only for generic values of the coupling $\alpha$, since after fixing diffeomorphism invariance, all the remaining modes of the graviton do propagate. In our case, since we are considering black holes at the non-generic, highly symmetric, Chern-Simons point, namely $\alpha\Lambda=-3/4$, the gravitational perturbations may be ill-behaved, since some of the components of the curvature of the background geometries are a constant such that not all the components of the graviton perturbations will appear in the linearized field equations after fixing diffeomorphism invariance. Therefore, at the Chern-Simons point the linearized analysis is singular on these types of background (the same occurs with the globally AdS solution or with the dimensionally continued black hole \cite{Banados:1993ur}), and in order to study perturbative stability one must go beyond the linearized regime, studying the evolution of initial data with energy bounded from above, and then evaluating whether the evolution leads to a spacetime that is close, in some functional sense, to the original background. Such analysis is beyond the scope of our work, and actually it has never been addressed in the literature, even though it is a fundamental problem of mathematical-physics in the context of these theories. Considering these issues, we have used the scalar field as a probe to explore the dynamics of a field of spin-0 (instead of spin-2) on top of the new rotating and non-rotating backgrounds we have constructed.

Finally, we constructed rotating solutions with non-vanishing torsion. These solutions are determined by four integration constants, one of them controlling the intensity of the contorsion. The spacetime metric in this case is given as a warped product of a hyperbolic two-dimensional space and a three-dimensional rotating spacetime with constant Ricci scalar. The torsion ansatz is compatible with the isometries of the latter. These results define a precursor for the embedding of our solutions in Chern-Simons AdS-supergravity. For example, the $SU(2, 2|N )$ CS-supergravity with $\mathcal{N}=4$ supersymmetries has shown to be a fruitful playground for the construction of static BPS, charged solutions. It would be natural to consider the embedding of the spacetimes constructed in the present work, now in the context of the CS-supergravity theory, since even in standard gauged SUGRA, charged gravitational solitons have given rise to new BPS solutions with interesting properties (see e.g. \cite{Anabalon:2021tua}, \cite{Anabalon:2020loe} and \cite{Aguayo:2022ydj}). We expect to explore some of these lines in the future. 

\section*{Acknowledgements}

We thank Eloy Ayon-Beato and Gaston Giribet for enlightening comments. L.T. appreciates the support of Dirección de Postgrado UdeC and FONDECYT grant 1221504. M.A. thanks the support of ANID Fellowship 22221854. The work of A.A. is supported in part by the FONDECYT grants 1200986, 1210635, 1221504, 1230853 and 1242043. The work of D.A. is supported in part by the FONDECYT grant 1242043. F.I. is supported by FONDECYT grant 1211219 from the Government
of Chile. N.E.G is partially supported by CONICET grant PIP-2023-11220220100262CO and UNLP grant 2022-11/X931. J.O. is partially supported by FONDECYT Grant 1221504. C. Quinzacara is supported by
FONDECYT grant 11231238.

\bibliographystyle{elsarticle-harv}


\end{document}